\documentclass{iau}
\usepackage{natbib}
\usepackage{graphicx}
\usepackage{aas_macros}

\newcommand{\jcap}{\textit{JCAP}}
\newcommand{\mnrasiau}{\textit{MNRAS}}
\newcommand{\prdiau}{\textit{Phys.~Rev.~D.}}
\newcommand{\apjiau}{\textit{ApJ}}
\newcommand{\apjliau}{\textit{ApJL}}
\newcommand{\aapiau}{\textit{A\&A}}

\title[The intrinsic alignment of galaxies] 
{Large-scale structure and the intrinsic alignment of galaxies}

\author[Jonathan Blazek, Uro\v{s} Seljak \& Rachel Mandelbaum]   
{Jonathan Blazek$^{1, \thanks{email: {\tt blazek@berkeley.edu}}}$,
Uro\v{s} Seljak$^{2}$,
\and Rachel Mandelbaum$^{3}$}

\affiliation{$^1$Center for Cosmology and AstroParticle Physics, Department of Physics,\\Ohio State University, Columbus, USA \\[\affilskip]
$^2$Departments of Physics and Astronomy and Lawrence Berkeley National Laboratory,\\University of California, Berkeley, USA \\[\affilskip] 
$^3$McWilliams Center for Cosmology, Department of Physics, Carnegie Mellon University, Pittsburgh, PA, USA}

\pubyear{2014}
\volume{308}  
\jname{The Zeldovich Universe: Genesis and Growth of the Cosmic Web}
\editors{R. van de Weygaert, S. Shandarin, E. Saar \& J. Einsasto, eds.}
\begin{document}

\maketitle

\begin{abstract}
Coherent alignments of galaxy shapes, often called ``intrinsic alignments'' (IA), are the most significant source of astrophysical uncertainty in weak lensing measurements. We develop the tidal alignment model of IA and demonstrate its success in describing observational data. We also describe a technique to separate IA from galaxy-galaxy lensing measurements. Applying this technique to luminous red galaxy lenses in the Sloan Digital Sky Survey, we constrain potential IA contamination from associated sources to be below a few percent.
\end{abstract}

\firstsection 
\section{Introduction}

Coherent, large-scale correlations of the intrinsic shapes and orientations of galaxies are a potentially significant source of systematic error in gravitational lensing studies, with the corresponding potential to bias or degrade lensing science results. This intrinsic alignment (IA) of galaxies has been examined through observations \citep[e.g.][]{hirata07,joachimi11}, analytic modeling \citep[e.g.][]{catelan01, hirata04}, and simulations \citep[e.g.][]{schneider12, tenneti14a} - see \citet{troxel14rev} for a recent review. As our understanding has improved, IA has also emerged as a potential probe of large-scale structure as well as halo and galaxy formation and evolution \citep[e.g.][]{chisari13}.

We present recent developments in understanding IA. We first describe analytic modeling of IA, focusing on the tidal alignment model (sometimes called the ``linear alignment model''). This model is tested and developed to consistently include nonlinear effects \citep{blazek11,blazek15}. We then present a technique to separate the IA and weak lensing signals in a galaxy-galaxy lensing measurement \citep{blazek12}.

\section{Tidal alignment}

{\underline{\it Model predictions and comparison to data}}. The tidal alignment model \citep{catelan01, hirata04} posits that the intrinsic ellipticity components of a galaxy, denoted $\gamma_{(+,\times)}$ are proportional to the tidal field:
\begin{equation}
\gamma^I_{(+,\times)}=-\frac{C_1}{4\pi G}(\nabla_x^2-\nabla_y^2,2\nabla_x\nabla_y)S[\Psi(z_{\rm IA})],
\end{equation}
where $C_1$ parameterizes the strength of the alignment, $\Psi$ is the gravitational potential, and $S$ is a filter that smooths fluctuations on halo or galactic scales.\footnote{The proper form and scale of this smoothing can have an important impact on model predictions on small scales. It is the subject of ongoing work \citep{chisari13,blazek15}.} This type of alignment is most likely to arise in elliptical galaxies for which angular momentum does not play a significant role in determining shape and orientation. For spiral galaxies, the acquisition of angular momentum, for instance through ``tidal torquing,'' generally leads to quadratic dependence on the tidal field \citep[e.g.][]{hirata04}. The tidal alignment contribution is the lowest-order function of the gravitational potential with the necessary symmetry and is thus expected to dominate IA correlations on sufficiently large scales. The redshift at which the alignment is set, $z_{\rm IA}$, is determined by the astrophysical processes involved in galaxy formation and evolution. It is sometimes assumed that $z_{\rm IA}$ is during matter domination when a halo first forms. However, late-time accretion and mergers could have a significant impact on IA, in which case the relevant $z_{\rm IA}$ could be closer to the observed redshift, allowing for significant nonlinear evolution of the tidal field \citep{blazek15}.

\begin{figure}[t!]
\begin{center}
\resizebox{\hsize}{!}{
\includegraphics{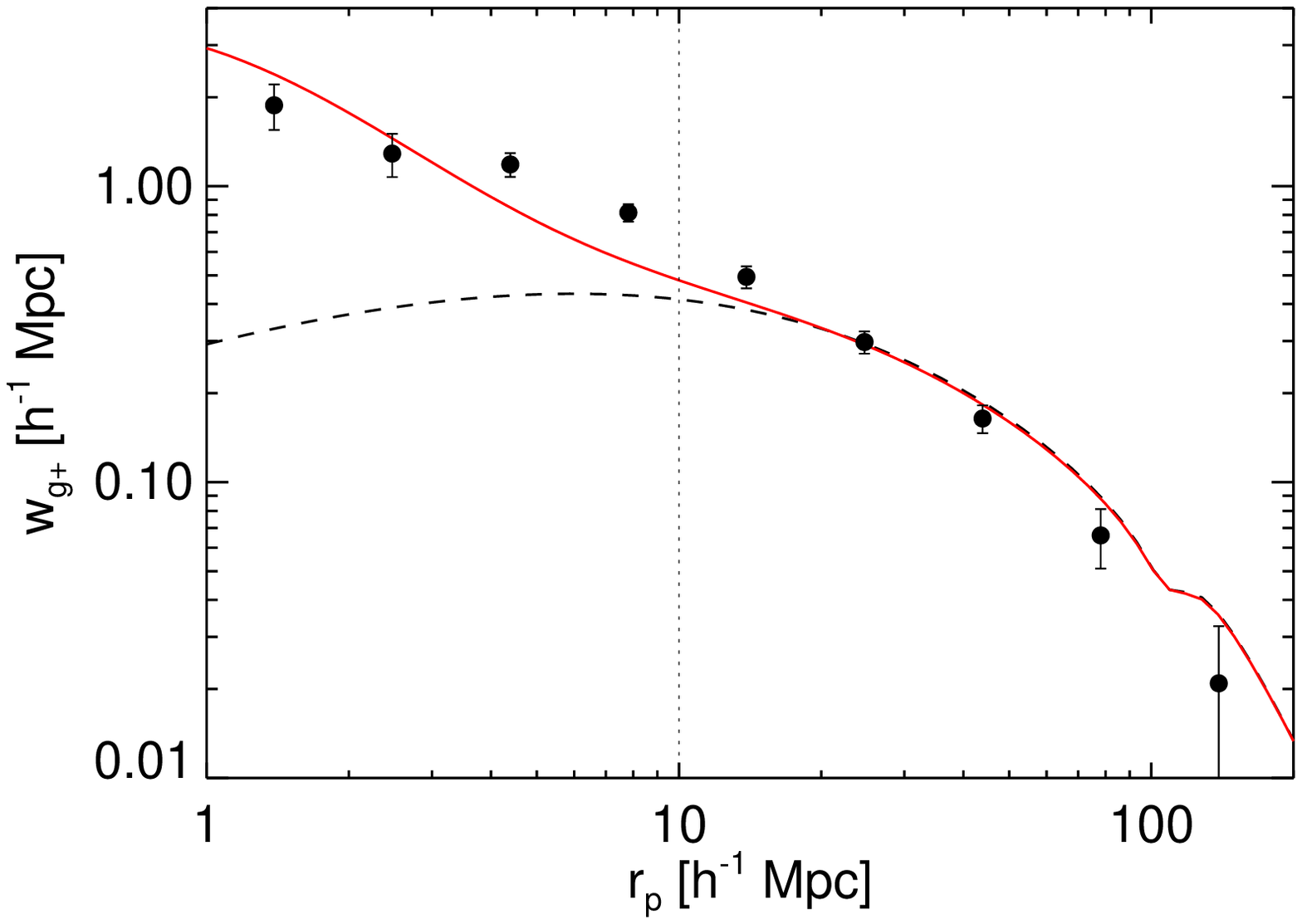}
\includegraphics{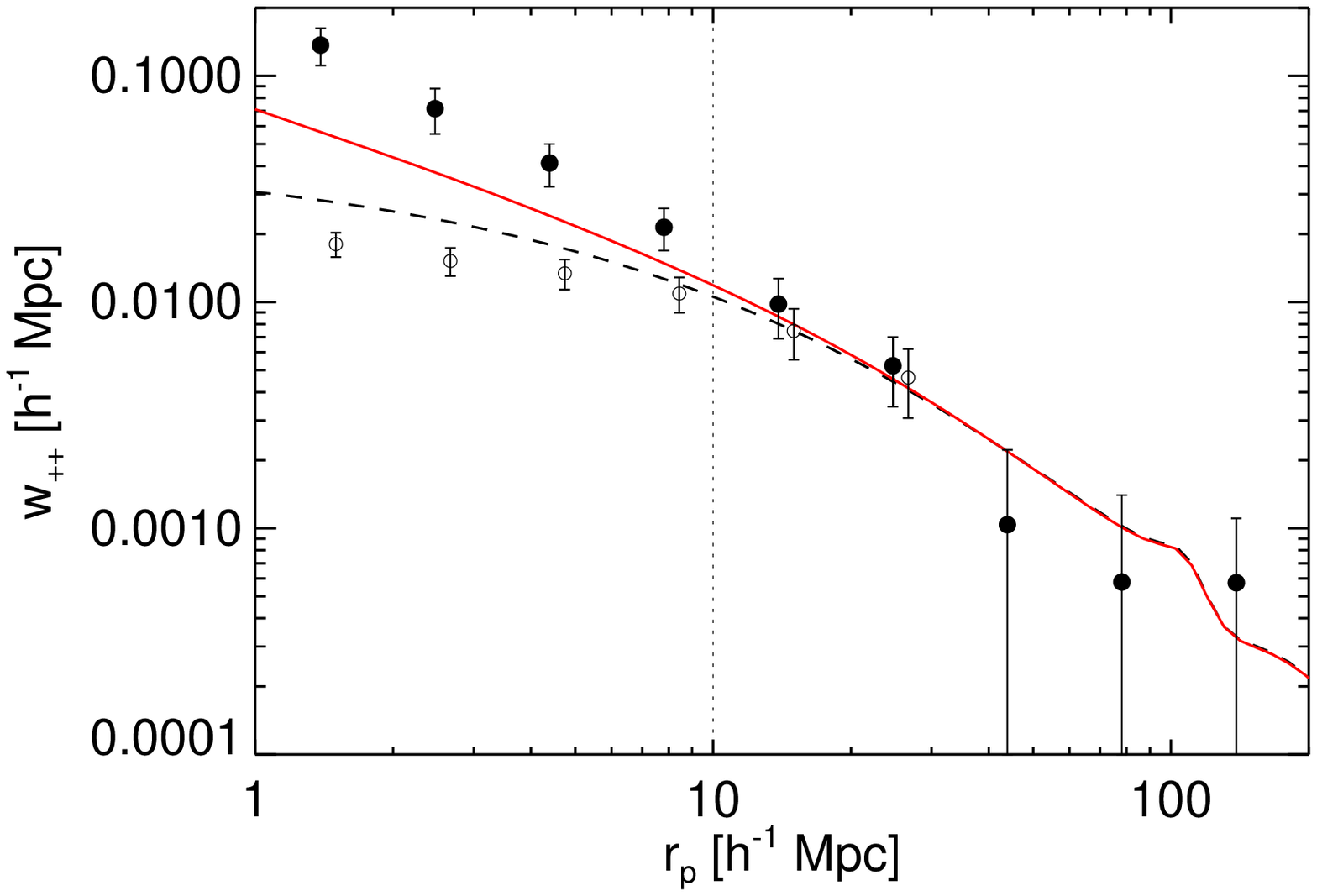}
}
\end{center}
\caption{The projected correlation function between galaxy position and shape $w_{g+}$ ({\it left panel}) and projected shape autocorrelation $w_{++}$ ({\it right panel}) are shown as a function of projected separation, $r_p$. Data points show the measurements of \cite{okumura09a,okumura09b}. The $w_{++}$ measurements have been projected along the line-of-sight, and open circles indicate the original measurement without a correction required to account for source clustering - see \cite{blazek11} for more information. Dashed lines indicate the linear order prediction of the tidal alignment model, while solid lines include the Halofit correction for nonlinear dark matter clustering \citep{smith03}.}
\label{fig:corr_functs}
\end{figure}

To test the predictions of the tidal alignment model, we consider measurements of the auto- and cross-correlations between the shapes and positions of luminous red galaxies (LRGs) in the Sloan Digital Sky Survey \citep[SDSS;][]{okumura09a,okumura09b}. Since they are both highly biased and elliptical, LRGs should exhibit strong alignment in agreement with the tidal alignment model. Indeed, as seen in Fig.~\ref{fig:corr_functs}, the tidal alignment model provides a good description of correlations at large projected separations ($r_p \gtrsim 10~h^{-1}{\rm Mpc}$). Fitting the model on large scales to both auto- and cross-correlations yields a consistent value of $C_1 \rho_{\rm crit} \approx 0.12 \pm 0.01$ \citep{blazek11}. Agreement with measurements on smaller scales is improved when the nonlinear evolution of dark matter clustering is included, producing what is sometimes called the ``nonlinear alignment'' (NLA) model \citep{bridle07}.

{\underline{\it Consistent treatment of nonlinear effects}}. Four effects can produce nonlinearities in the intrinsic shape correlations: (1) nonlinear dependence of intrinsic galaxy shape on the tidal field (e.g., quadratic tidal torquing); (2) nonlinear evolution of the dark matter density field, leading to nonlinear evolution in the tidal field; (3) a nonlinear bias relationship between the galaxy and dark matter density fields; (4) the IA field actually observed is weighted by the local density of galaxies used to trace the shapes. In the tidal alignment model, intrinsic galaxy shapes depend linearly on the tidal field, even on small scales. However, nonlinearities from the other three effects must still be considered. The NLA model includes the nonlinear evolution of the dark matter density but does not consider other nonlinear contributions. Thus, although the NLA approach improves the model fit to data, it is not fully consistent and omits important astrophysical effects. Instead, we examine all nonlinear contributions in the context of tidal alignment \citep{blazek15}, including terms that contribute at next-to-leading order while simultaneously smoothing the tidal field (e.g. at the Lagrangian radius of the host halo). We find that the effects of weighting by the local galaxy density can be larger than the correction from the nonlinear evolution of dark matter density, especially in the case of highly biased tracers, since the leading effect from weighting scales with the linear galaxy bias. Contributions from nonlinear galaxy bias are also appreciable, although subdominant.

\section{Separating IA from galaxy-galaxy lensing}

\begin{figure}[t!]
\begin{center}
\resizebox{\hsize}{!}{
\includegraphics{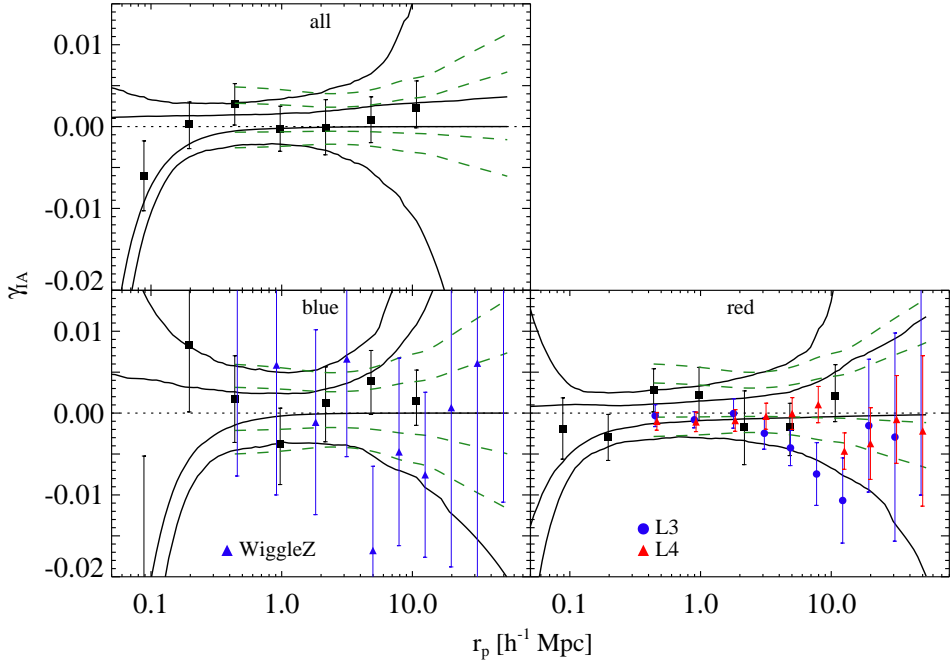}
}
\end{center}
\caption{Constraints on IA contamination to the measured shear for each physically associated source (all, blue, and red) are shown. Data points with errors are model-independent constraints. Solid (dashed) lines denote 68\% and 95\% confidence intervals assuming a power-law (LRG observational) model. Previous spectroscopic results for SDSS red galaxies from \cite{hirata07} and for WiggleZ galaxies from \cite{mandelbaum11} are shown for comparison.}
\label{fig:IAgglens}
\end{figure}

In galaxy-galaxy lensing, the shapes of background galaxies are used to probe the dark matter distribution around lens objects, often by stacking images at lens centers in order to increase the signal. Scatter in the photometric redshifts for lensing sources \citep[e.g.][]{nakajima12}) can cause objects that are physically associated with the lens to be assigned a location significantly behind the lens, contaminating the lensing signal with shape correlations between physically associated sources and lens positions. Since the lensing signal measures the projected density profile of the lens objects, it should not depend on the redshift distribution of the background sources, assuming that the distribution is well known. However, as more distant sources (as defined by photometric redshift) are used, the fraction of physically associated objects and the corresponding IA contamination will  decrease. Thus, by dividing the lens-source pairs into tomographic bins (by redshift separation), we are able to isolate the lensing signal from the IA signal. We apply this method to LRG lenses \citep{kazin10} using the SDSS lensing catalog described in \cite{reyes12} to place constraints on the average IA contamination per physically associated source \citep{blazek12}. Results are shown in Fig.~\ref{fig:IAgglens} and compared with IA measured directly in spectroscopic samples. These constraints correspond to a maximum IA contamination of a few percent at projected separations of $0.1 < r_p <10~h^{-1}{\rm Mpc}$. The tighter constraints for red sources is due to their stronger clustering rather than an underlying trend in IA amplitude.

\section{Conclusions}

Continued progress in understanding IA is critical for current and upcoming lensing experiments. We have presented important developments in modeling IA through the tidal alignment model, including a consistent treatment of nonlinear effects. Combined with results from simulations and techniques for predicting IA in the deeply nonlinear regime (e.g., a halo model approach), this improved tidal alignment model will allow more effective mitigation of IA contamination. We have also developed and applied a method to separate IA from the galaxy-galaxy lensing signal, providing both a clean lensing measurement and a probe of IA in galaxy samples without large numbers of spectroscopic redshifts. This technique should be an important tool in future measurements.

\clearpage

\end{document}